\documentclass[a4paper,10pt]{article}

\usepackage{graphicx,amsmath,amssymb}
\usepackage{bm}
\usepackage{color}
\usepackage{setspace,textcomp}
\usepackage{authblk}

\begin{document}
 
\title{On the origin of burstiness in human behavior: The wikipedia edits case}
\author[1]{Yerali Gandica}
\author[2]{Joao Carvalho}
\author[2]{Fernando Sampaio Dos Aidos}
\author[1]{Renaud Lambiotte}
\author[1]{Timoteo Carletti}
\affil[1]{Department of Mathematics and Namur Center for Complex Systems - naXys, University of Namur, rempart de la Vierge 8, B 5000 Namur, Belgium}
\affil[2]{{CFisUC}, Department of Physics, University of Coimbra, 3004-516 Coimbra, Portugal}

\date{\today}

\maketitle
\singlespacing
\begin{abstract}
A number of human activities exhibit a bursty pattern, namely periods of very high activity that are followed by rest periods. 
Records of this process generate time series of events whose inter-event times follow a probability distribution that 
displays a fat tail. The grounds for such phenomenon are not yet clearly understood. In the present work we use the freely
available Wikipedia's editing records to tackle this question by measuring the level of burstiness, as well as the memory 
effect of the editing tasks performed by different editors in different pages. Our main finding is that, even though the 
editing activity is conditioned by the circadian 24 hour cycle, the conditional probability of an activity of a given 
duration at a given time of the day is independent from the latter. This suggests that the human activity seems to be 
related to the high ``cost" of starting an action as opposed to the much lower ``cost" of continuing that action.
\end{abstract}
\newpage \baselineskip1.0cm
\singlespacing
\section{Introduction}

The digital media is an important component of our lives.
Nowadays, digital records of human activity of different sorts are
systematically stored and made accessible for academic research. 
Hence a huge amount of data became available on the past couple of decades,
which allows for a quantitative study of human behaviour.  
For a long time, scholars from different backgrounds have been studying this
field: a more general approach by
sociologists, and some specific studies by economists and social psychologists. 
However, some very interesting - and elemental - properties have remained
outside the grasp of the research studies for lack of reliable massive stored data. 
The increasing amount of data that is being gathered in this digital age is
progressively opening up new possibilities for quantitative studies of these
features.  
One such aspect, detected by means of data-gathering, is human bursty
behaviour, that is the mankind activity characterised by intervals of rapidly
occurring events separated by long periods of inactivity~\cite{barabasi2005}.  
The dynamics of a wide range of systems in nature displays such a behaviour \cite{hidalgo}.

Given the highly non-linear nature of human actions, their study could
hence benefit from the insights provided by the field of complex systems. For
the human being, the bursty behaviour phenomenon has been found to modulate
several activities, such as sending letters, writing email messages, sending mobile SMS,
making phone calls and web browsing
\cite{barabasi2006,barabasi2008,Schellnhuber,amaral2009,kartez2012}.  
The first works in this field suggested a
decision-based queuing process, according to which the next task to be
executed is to be chosen from a queue with a hierarchy of importance, in order 
to explain this behaviour. 
Different kinds of hierarchies were tested, such as taking into account the
task length and deadline constraints
\cite{barabasi2005,barabasi2006,barabasi2008}. Later on, Malmgren
{\it et al.} \cite{amaral2008,amaral2009} argued that decision making is not a
necessary component of the bursty human activity patterns. Instead, 
they maintained that this feature is caused by cyclic constraints in life and
they proposed a mechanism based on the coupling of a cascading activity to
cyclic repetition in order to explain it. Nonetheless, recently, Hang-Hyun Jo
{\it et al.} \cite{kartez2012} applied a de-seasoning method to remove the
circadian cycle and weekly patterns from the time series, and obtained similar 
inter-event distributions, before and after this filtering procedure. 
In this way, the authors concluded that cyclic activity is also not a necessary ingredient of bursty behaviour. 
However they do not provide an explanation for the fundamental origin of burstiness. 
Such explanation is still missing in the literature and the scope of this article is
to add some new insights into this issue by looking into Wikipedia editing. 

The success of research in digital social patterns hinges on the access to high quality data. 
Even though the availability of recorded data and its accessibility are
rapidly increasing, many data sets are not freely available for
research. Wikipedia (WP) is an important exception, as not only is it considered a robust and trustworthy source of information
\cite{giles}, but it is also easily accessible via the API \cite{wikiAPI} or
the different available dumps \cite{wikidumps}. The accessible data contain the whole
editing history record for both pages and 
editors.
We start by showing that
this database captures the bursty nature of the human actions, and that the human activity region described
by the bursty coefficient and memory effect in \cite{barabasi2008} must be increased if it is to accommodate the editing
activity of Wikipedia. 
We also show the cyclic nature of the data and the circadian patterns in WP editing. Most interestingly, we show how, 
although the editing activity is strongly constrained by the time of day at which it happens, 
the final probability distribution of the inter-event times is quite independent of that, 
displaying a fat tail that is robust against changes of editing time during the day. 
This suggests that burstiness in human behaviour is mostly independent of cycles in life.
Finally, we look into the effect that stationarity may have on burstiness; whether a change on editing activity over time may be the cause for the
fat tail of the inter-event distribution. Our results indicate that burstiness is common to all editors, whether their editing history is stationary 
or not, which suggests that there does not seem to be a cause and effect relationship between non-stationarity and burstiness. 
Having discarded cycles in life and non-stationarity as possible causes for burstiness, 
our findings suggest a sort of robust distribution for the time between activities, set internally by individuals.

\section{Methods}

Our data sample is a database of Wikipedia (WP) edits of all the pages written in English
in the period of $10$ years ending on January $2010$; the dump contains $4.64\times 10^6$ 
pages \cite{data}. For each entry, we know: the Wikipedia page name, the
edit time stamp and the identification of the editor who did the changes. We 
analyse the information separately for each page and for each editor and, in order to
reduce the impact of outliers, we did not consider the pages or editors
with less than $2000$ activities. This number is a good compromise between having
enough pages/editors and the pages being frequently updated or the editor
being reasonably active in this time span. We also remove from the data the edits
made by WP-bots, which are
programs that go through the WP carrying out automatically repetitive and
mundane tasks to maintain the WP pages (as software programs, their edit
pattern is different from the humans'). A list of the bots can be found in \cite{wikibots}. Moreover, we discard entries 
associated to IP's and only consider editors who login before editing, so that the editor is univocally identified.

In order to obtain the power spectrum of the editing activity shown in Fig. \ref{fourier1} for editors and in Fig. 
\ref{fourier2} for pages, we binarized our activity record for any editor and any page with small enough bins to 
contain at most one edit. Finally we performed a fast Fourier transform (FFT) on such preprocessed data. Although in every 
5,000 edits there is one second with more than one edit, we checked that the results do not change if these values are 
truncated to one (for the editors this must correspond to a double saving operation in the same second and this should 
reasonably be counted as one).

\section{Results}

\subsection{Bursty level and Memory effects}

Let $E_i$ be a variable that identifies each editor in our sample, $t_n^{E_i}$ the time at which $E_i$ edits for the nth 
time and $\tau_n^{E_i}=t_{n+1}^{E_i}-t_n^{E_i}$ the nth inter-event time for that editor (where obviously $n\ge 1$). 
For the sake of clarity in the equations, we avoid this cumbersome notation and omit the superscript $E_i$, writing just 
$t_n$ and $\tau_n$ for short. These should be understood as referring the specific editor that is being considered. 
The same will apply to the other variables in the following equations.

A standard measure of burstiness has been provided by Goh {\it et al.} \cite{barabasi2008}. For editor $E_i$ we obtain:
\begin{equation}
B_i \equiv \frac{\sigma - m}{\sigma + m}\,
, 
\label{eq1}
\end{equation}
where $m$ and $\sigma$ are, respectively, the mean and
standard deviation of the set of values $\tau_n$ which refer to editor $E_i$. Parameter $B_i$
measures the magnitude of the variance in 
comparison to the mean, and goes from -1 ($\sigma \ll m$), for periodic
signals, to 1 ($\sigma \gg m $) for strongly bursty distributions.

In order to have a better characterisation for the general behaviour of the
editing process, we also compute the memory coefficient $M_i$, for editor $E_i$, which is also defined
in \cite{barabasi2008}: 
\begin{equation}
M_i \equiv  \frac{1}{n_{\tau} -1} \sum_{j=1}^{n_{\tau} -1}
\frac{({\tau}_j - m_{\tau})({\tau}_{j+1} - m_{\tau}^{\prime})}{
  \sigma_{\tau} \sigma_{\tau}^{\prime}}\, ,
\label{eq2}
\end{equation}
where $n_{\tau}$ is the number of inter-event values for that particular
editor (equal to the number of edits minus 1), and $m_{\tau}$ and
$\sigma_{\tau}$ ($m_{\tau}^{\prime}$ and $\sigma_{\tau}^{\prime}$) are,
respectively, the average and the standard deviation of the set of all
inter-event times except the last one (the first one).  
This coefficient is very similar to the autocorrelation function for
consecutive inter-events and is positive when short (large) inter-event
times tend to be followed by short (large) inter-event times, negative in the
opposite case and otherwise close to zero.   

\begin{figure}[tbh]

\begin{center}
\scalebox{0.4}{\includegraphics{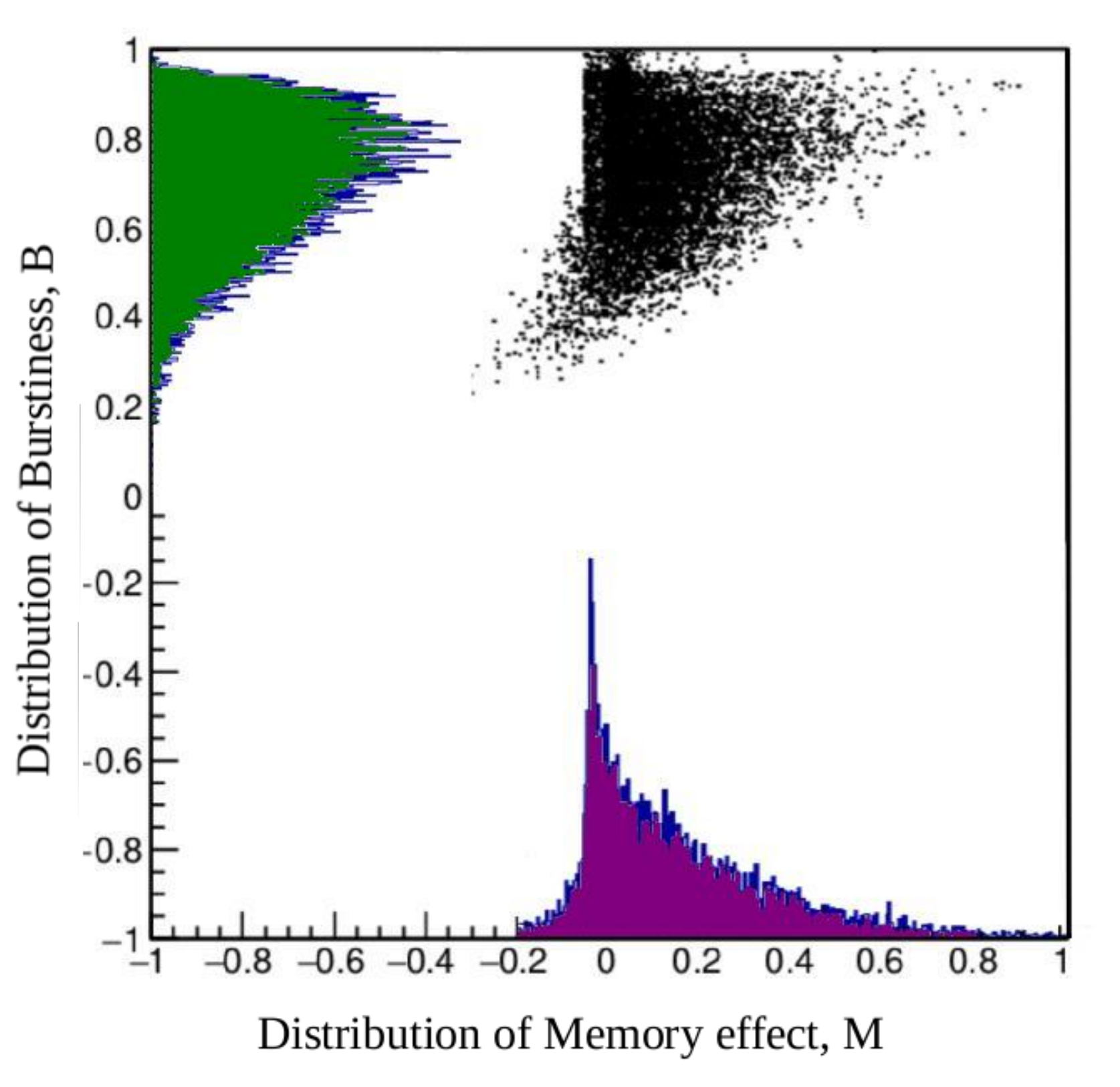}} \caption{Scatter plot for the
  joint distribution of Burstiness, $B_i$, versus Memory coefficient,
  $M_i$.}
\label{figbm}
\end{center}
\end{figure}

In Fig.~\ref{figbm} we show the scatter plot of the joint distributions of $M_i$ and $B_i$. We can see that the main population
has high values of burstiness and, at the same time, low memory effects, meaning that there is small correlation between
the sizes of the time span of two consecutive inter-events. The average values obtained for $B_i$ and $M_i$ are 
$\langle B\rangle=0.770$ and $\langle M\rangle=0.125$, values that result to be outside the region of human behaviour 
reported by Goh {\it et al.} \cite{barabasi2008} (red zone in figure 4 of  \cite{barabasi2008}). 

\begin{figure}[tbh]
\begin{center}
\scalebox{0.6}{\includegraphics{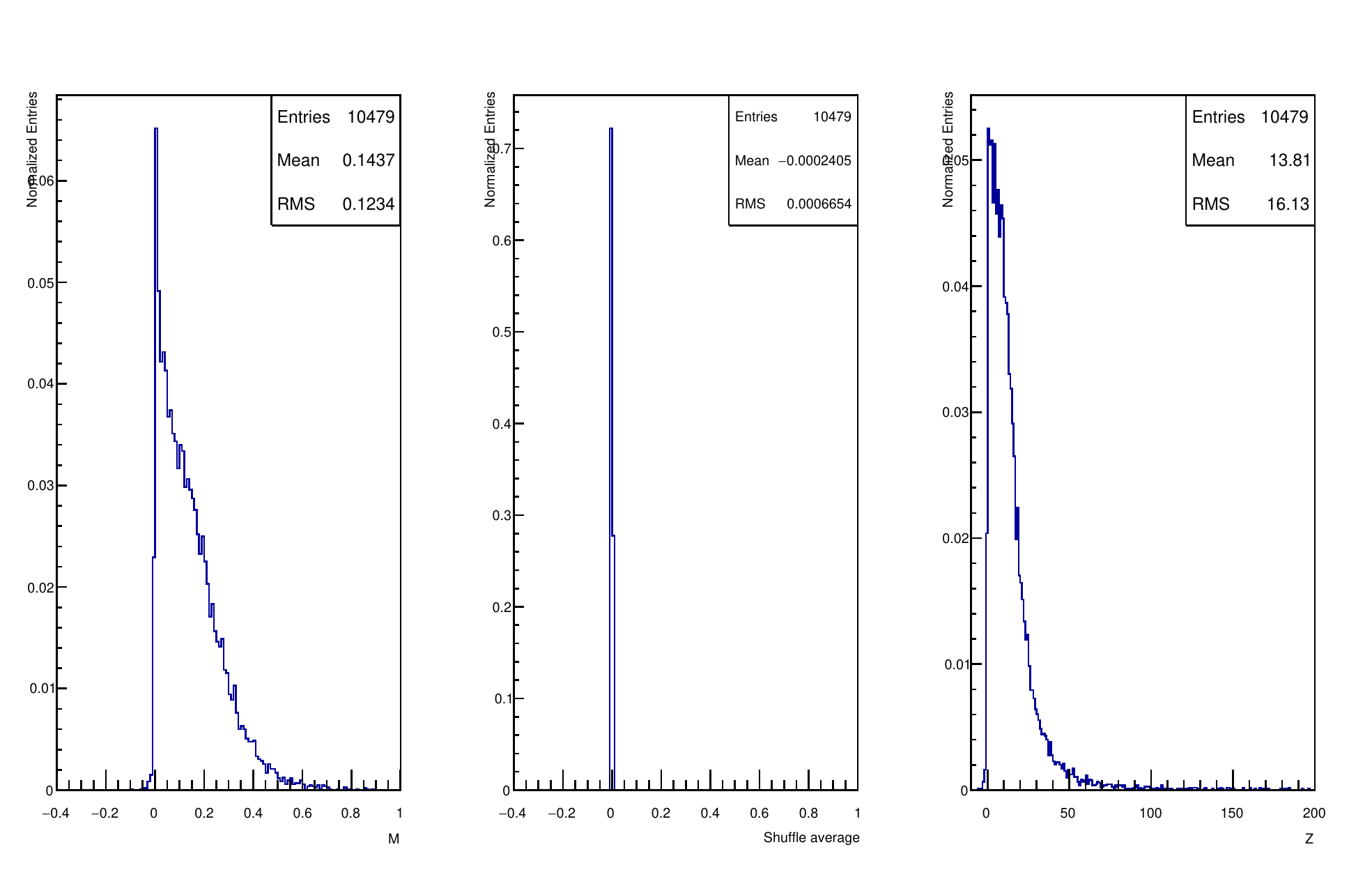}} \caption{The memory
coefficient. Left panel: Distribution of $M_i$ for all the editors with more than $2000$ edits. 
Middle panel: Distribution of the averages $\overline{M}_i$ for the null-model
computed using $100$ replicas for each editor. Right panel: Distribution of
the z-score values.} 
\label{figm}
\end{center}
\end{figure}

The value for $\langle M\rangle$ is small but it is statistically different from zero, as we now show. 
The normalised distribution of the values of $M_i$ for all the editors is shown in the left panel of
Fig.~\ref{figm}. The distribution is very skewed, with very few negative
values. To determine whether the average value obtained for $M_i$ is statistically relevant (different from zero)
we built a null-model. For each editor, we reshuffled the sequence of
inter-event times $\tau_i$ and obtained a new sequence $\theta_i$ where all
possible time correlations have been destroyed. Then, we computed the memory
coefficient $\tilde{M}_i$ of such new sequence. Finally we repeated such a
process $100$ times and computed the average $\overline{M}_i$ of all the memory coefficients,
as well as its standard deviation $\overline{\sigma}_i$. 

The normalised distribution of $\overline{M}_i$ is shown in the middle panel of
Fig~\ref{figm}. We can see that $\overline{M}_i$ is basically zero, due to the elimination of the memory 
(time correlations) by the reshuffling. On the right panel of the same figure we report the normalised
distribution of the z-scores $z_i=\frac{M_i-\overline{M}_i}{\overline{\sigma}_i}$.
From the latter plot we confirm
that the average of $\langle M\rangle-\overline{M}_i$ (which is basically equal to $\langle M\rangle$) is large enough, in terms of
standard deviations, to conclude that the value of $\langle M \rangle$ is statistically relevant and
different from zero.

\subsection{The cyclic nature of WP editing}

\begin{figure}[tbh]
\begin{center}
\scalebox{0.44}{\includegraphics{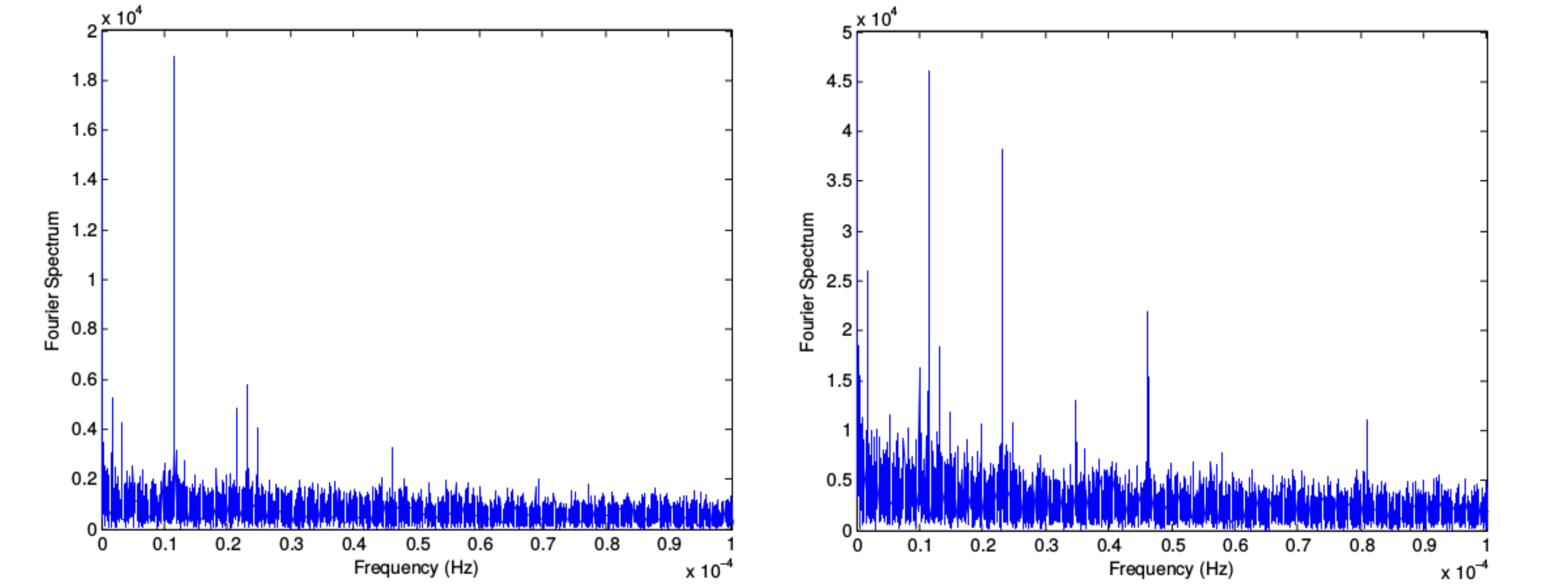}} \caption{Fourier spectrum of the
editing activity for editors $E_1$ and $E_2$. One can clearly identify the main peak $f_1\sim 1.157\times 10^{-5}$ Hz 
corresponding to a $24$ hours period and, especially on the right panel, several harmonics.}
\label{fourier1}
\end{center}
\end{figure}

WP editors, as much as anybody else, live in a dynamic environment, strongly
conditioned by periodic factors: day/night, weekdays/weekends,
seasons, etc. It it thus important to check whether their WP activity is also conditioned
by such rregularities. For this purpose, we selected the $100$ most active editors
and computed the Fourier power spectrum of each editor time activity, as explained in the methods. The results for two 
representative editors are reported in Fig. \ref{fourier1} where we show the power spectrum for 
editors $E_1$ (left) and $E_2$ (right). In each panel a main peak is clearly visible at $\sim 1.157\times
10^{-5}$ Hz corresponding to a period of $24$ hours as well as a second peak at $\sim 2.315\times
10^{-5}$ Hz associated to a $12$ hours period (a harmonic of the main peak). 
Other harmonics are also visible, especially in the right panel. We checked that these values are robust against the choice of the bin size, from a few
seconds up to several hours, used to perform the FFT. 
 
\begin{figure}[tbh]
\begin{center}
\scalebox{0.3}{\includegraphics{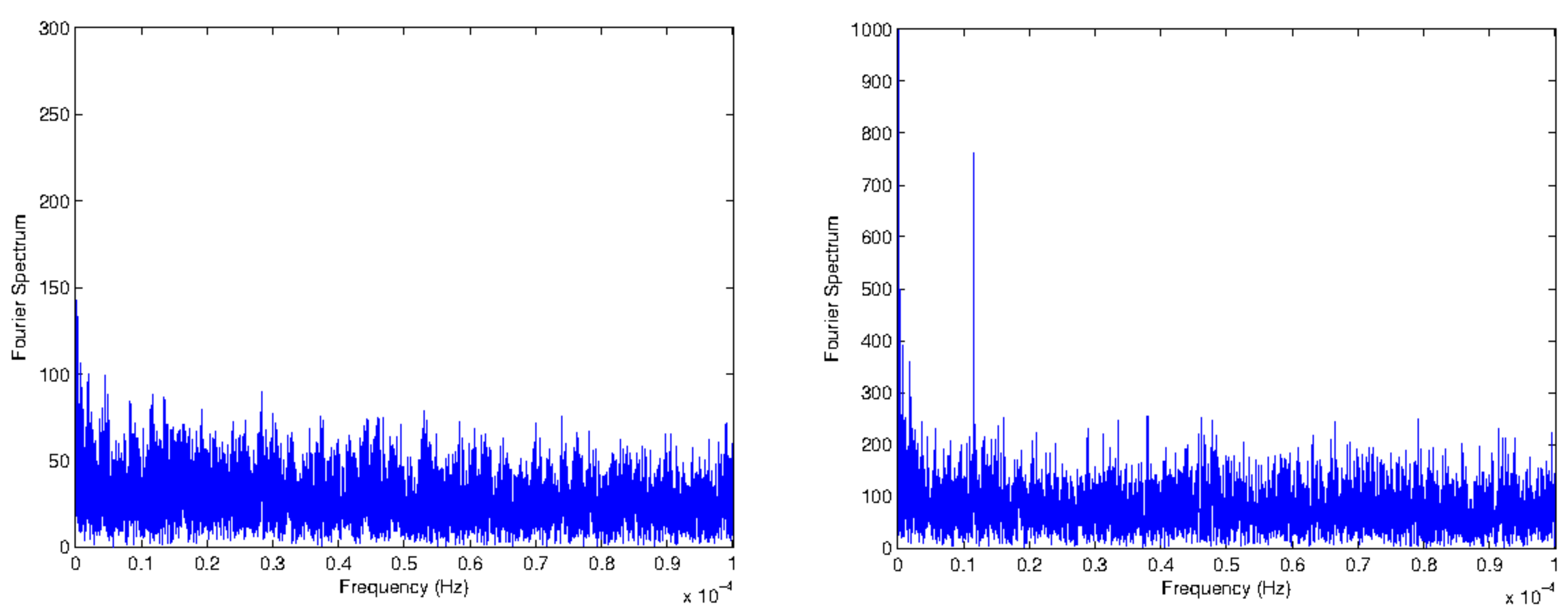}} \caption{Power spectrum of two different types of Wikipedia pages. 
The left panel corresponds to a generic issue and the right panel to a big company. Note the one day periodicity in the latter and the absence of peaks in the former. }
\label{fourier2}
\end{center}
\end{figure}

In order to check whether this periodicity is particular to human behaviour, we repeated this procedure for the WP pages.
We selected the 100 pages with more edits and evaluated the Fourier power spectrum in the same way that was done for 
the editors. We found that, in general, pages lack predominant peaks in the power spectrum. They appear only in pages 
related to records (for example births/deaths counting) or companies, which in general update their pages daily. In figure 
\ref{fourier2}, we show the power spectrum of a generic page in the left panel, as an example for the general behaviour 
found in the WP pages. In the right panel we show the power spectrum of the WP page of a big company. In the latter 
case, we can see clearly the one day period that is lacking in the generic WP page. It is interesting that pages do not 
show periodic behaviour, even though they are edited by editors who do show that behaviour. The conjugation of many 
different editors, with different editing rhythms, wipes out the periodicity, just like summing many oscillators with 
random phases.

\begin{figure}[tbh]
\begin{center}
\scalebox{0.5}{\includegraphics{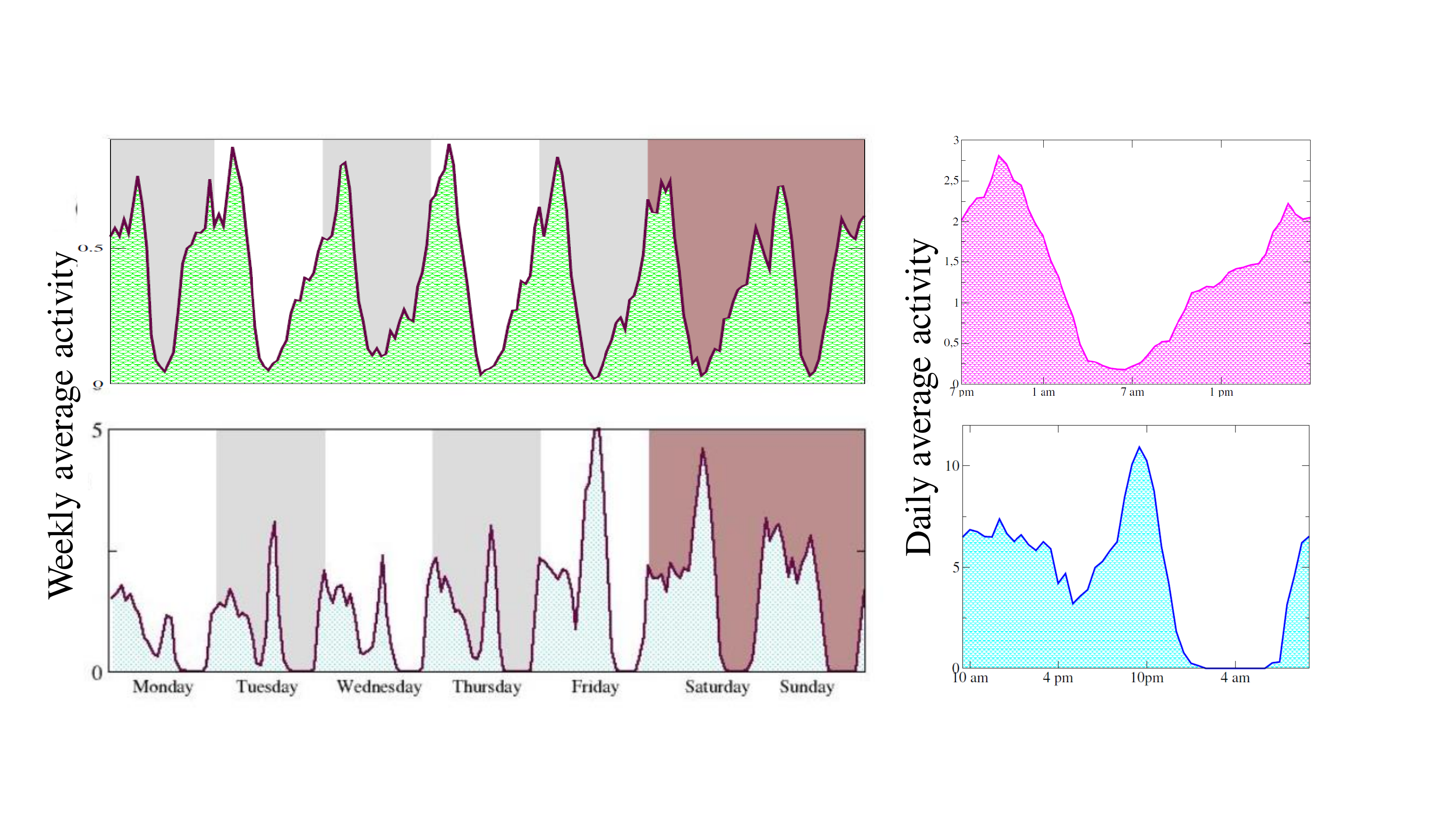}}
\end{center}
\caption{The left panels show the editing activity as a function of the week day, averaged over the 
data time span (about ten years) for two of the most active editors, $E_1$ (top) and $E_2$ (bottom). 
Circadian patterns are clearly visible. In general, we can identify a different editing activity over the week-ends (dark area). The 
right panel shows the average editing distribution along the day. One main peak
of activity is clearly visible, as well as a time span of very low activity (presumably corresponding to a resting period).}
\label{circadian}
\end{figure}

A better illustration of these circadian patterns can be found in Fig.
\ref{circadian}, which shows the average number of edits over a period of one week (left) and one day (right) of editors $E_1$ (top) 
and $E_2$ (bottom), as a function of the editing time. To build the daily (weekly) plot, we divided the whole time span of our
data of about ten years into days (weeks), and then divided each day (week) in 48 (168) windows with 30 (60) minutes each. For each window, we computed 
the average number of edits that were performed inside that window for every day (week) for the whole 10 year time span. That gives the daily (weekly) 
average activity which is plotted in Fig. \ref{circadian}.

The circadian patterns are visible in the plots; each editor exhibits clearly one peak (sometimes two peaks) of activity every day. 
We can also see a daily trough of low activity that will probably correspond to the time when the editor rests.

\subsection{Robustness of the inter-event distribution}

In the previous section we have shown that editing is strongly
influenced by the circadian cycle; in this section we analyse whether these circadian patterns have consequences on the inter-event probability 
distribution. To perform such an analysis we computed the probability distribution for
inter-events of a given duration assuming that they have been saved at a particular time of day. 
If this probability depends on the time of day at which they are recorded, 
then the time spent in a human activity is also dictated by cycles and thus the origin
of burstiness can possibly be ascribed to this phenomenon; on the
other hand, if the probability to perform an activity that lasts a definite amount of time does not
depend on the time of day at which it is registered, then we can conclude that burstiness in WP editing does not depend on
the periodically changing environment.

\begin{figure}[tbh]
\begin{center}
\scalebox{0.3}{\includegraphics{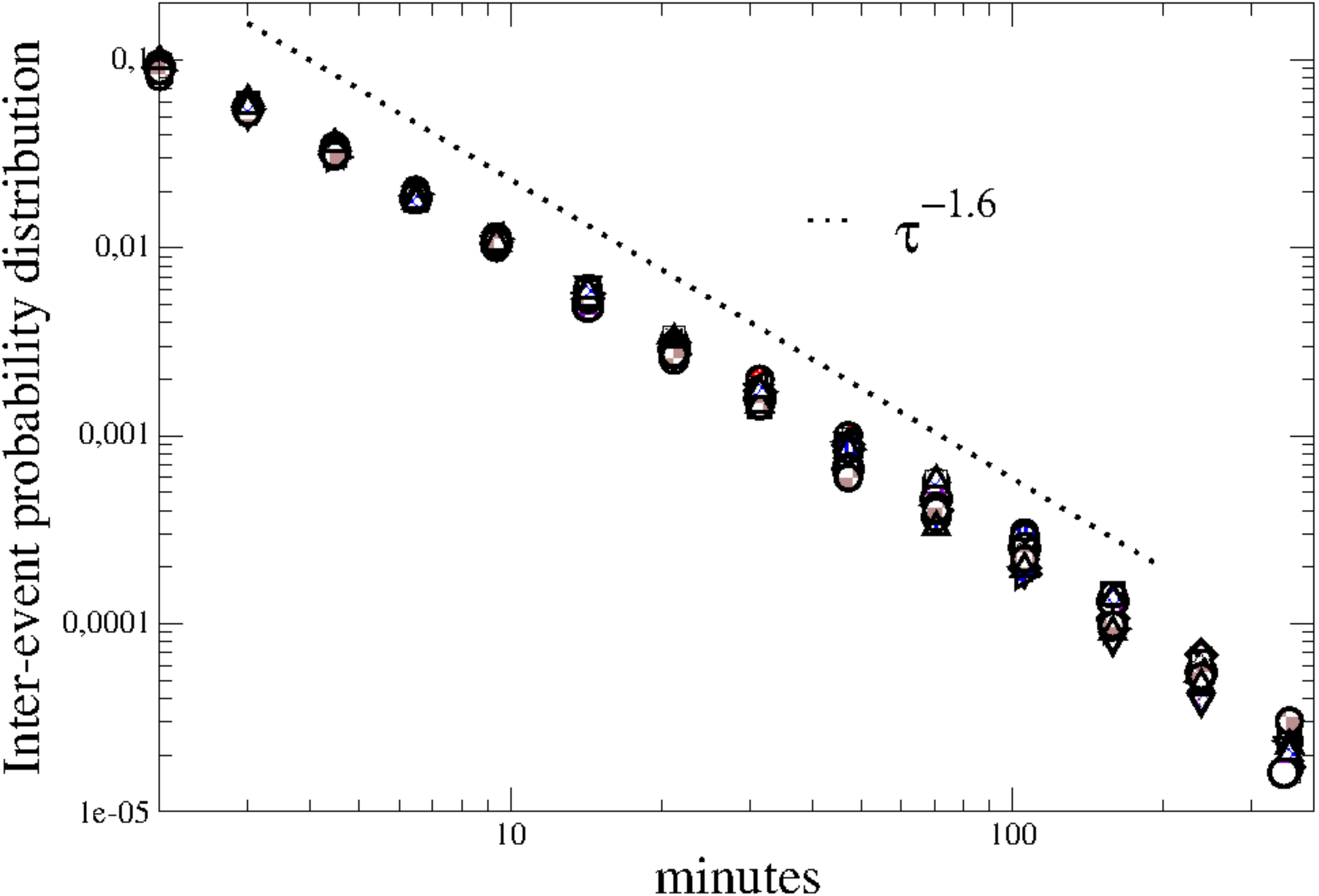} \includegraphics{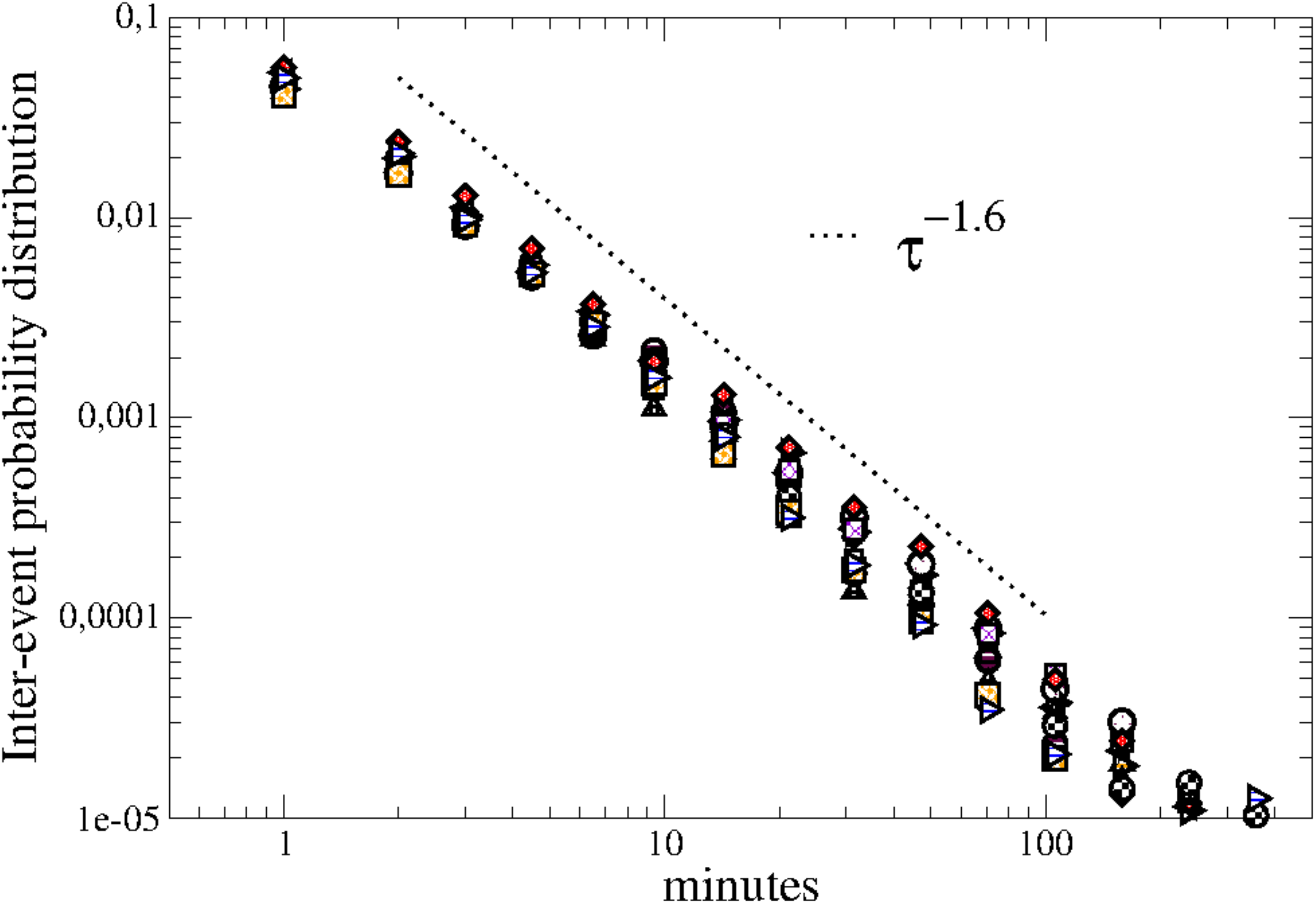}} \caption{Probability distributions for the inter-event times of
editors $E_1$ and $E_2$ for different values of the inter-event time. Each color-symbol represents the
probability distribution for the inter-event times that are recorded during each of the 17 one hour windows of the day. Seven 
windows were left out because the small quantity of data they contain was not sufficient to draw statistically 
sounding conclusions. One can clearly appreciate the similar fat-tail distributions of inter-events, independently of the 
circadian patterns of data. Observe also that the inter-event duration is limited to a few hours; this is due to our choice 
to present data for editors who have a very high activity and consequently relatively short inter-event durations.
In both panels, the dotted lines represent the best fit.} 
\label{interdiarydistri}
\end{center}
\end{figure}

Results reported in Fig. \ref{interdiarydistri} support the latter
hypothesis. The inter-event probability distributions computed in different
time windows, with a one hour time span, exhibit a similar fat tail when they are normalized by the number of event in that hour. 
Note that only 17 time windows are shown in Fig. \ref{interdiarydistri}. Seven windows were left out as they correspond to the 
periods of low activity shown in Fig. \ref{circadian} and had too little data to be relevant.

This seems to indicate that, although the event probability responds to strong
circadian patterns during the day, the bursty nature of the process is not a
consequence of that cyclic behaviour. In ref \cite{kartez2012}, the authors reported the independence of burstiness with 
respect to the 24 hour cycles, showing the same inter-event distribution in windows of one day. What we show here is the 
same inter-event distribution but in windows of one hour, where the brevity of the time windows provides a good proxy for 
the independence of the activity duration from the time of day of its realisation. This kind of universality has been 
reported in phenomena emerging from the joint action of several individuals \cite{y1,y2}. However, to the best of our 
knowledge, the occurrence of this phenomenon as a result of a single person's decision has not been reported previously in 
the literature. 

Finally we must discard any effects on the bursty nature of the data due to changes in the 
editing activity along the years~\cite{amaral2009}. We check the stationarity of the time series of editors by splitting
it into ten windows, each one corresponding to the same time span, and comparing the probability distributions of 
inter-events of all ten windows with each other. We have done that for the editing data of several editors.

\begin{figure}[tbh]

\begin{center}
\includegraphics[width=0.45 \textwidth]{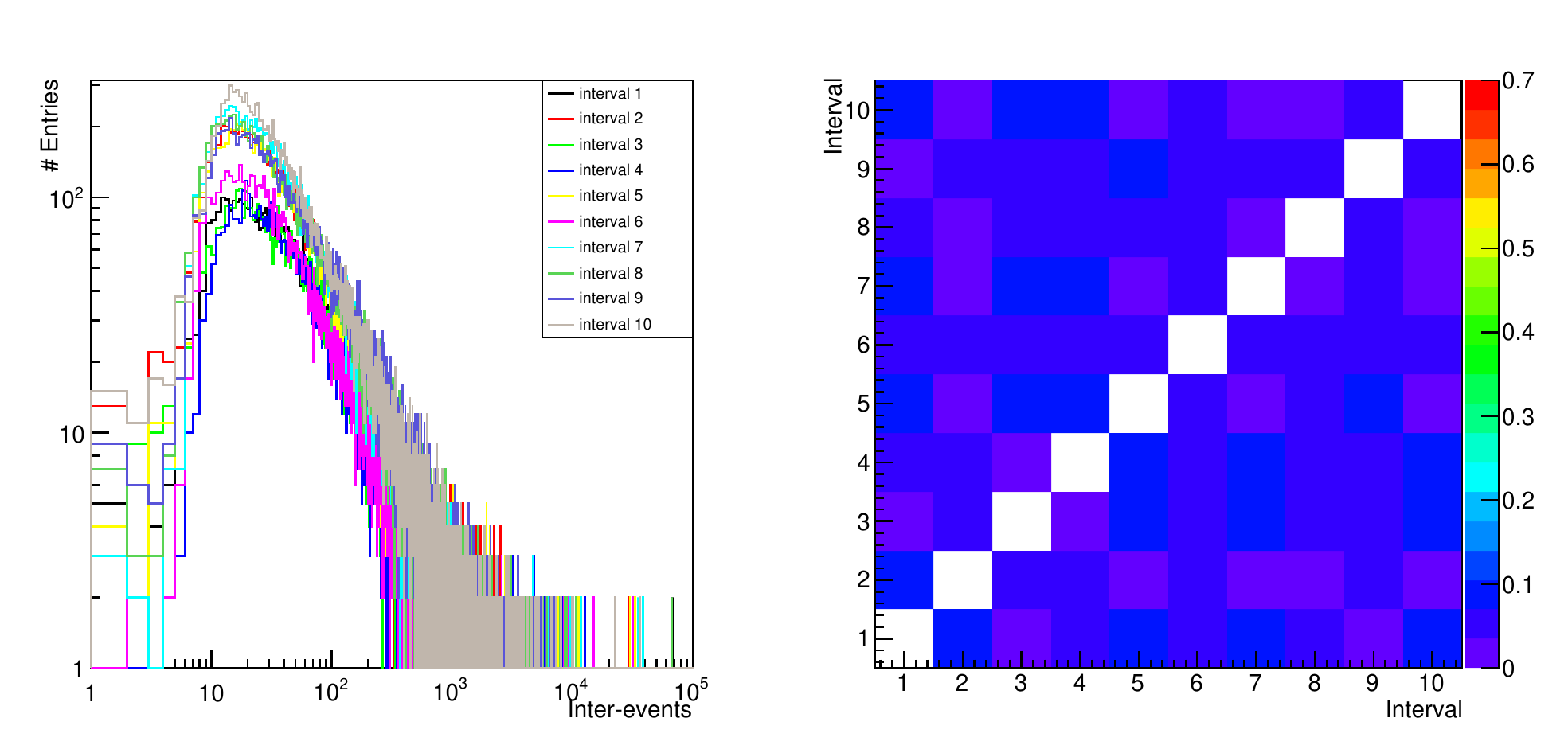}
\includegraphics[width=0.45 \textwidth]{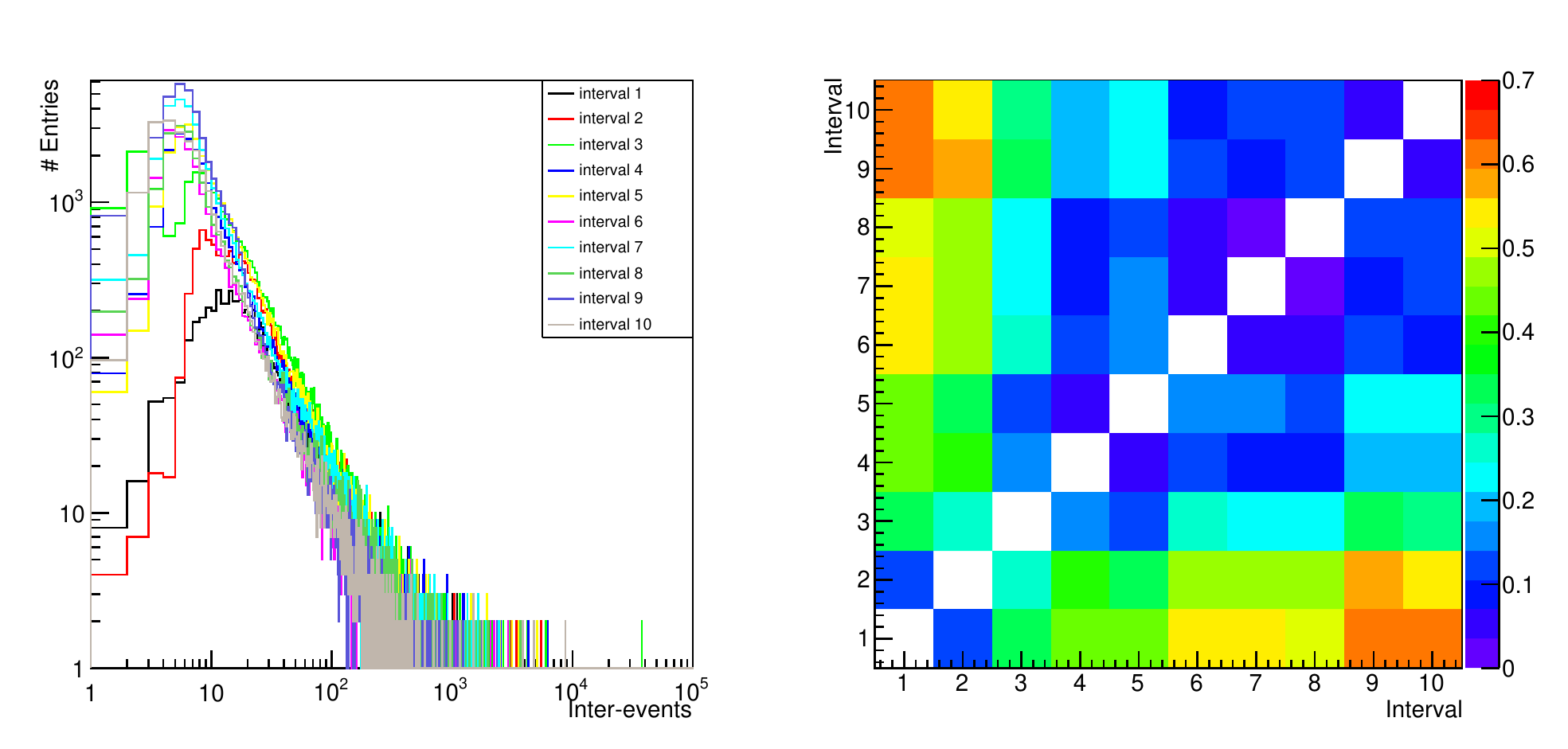}
\includegraphics[width=0.45 \textwidth]{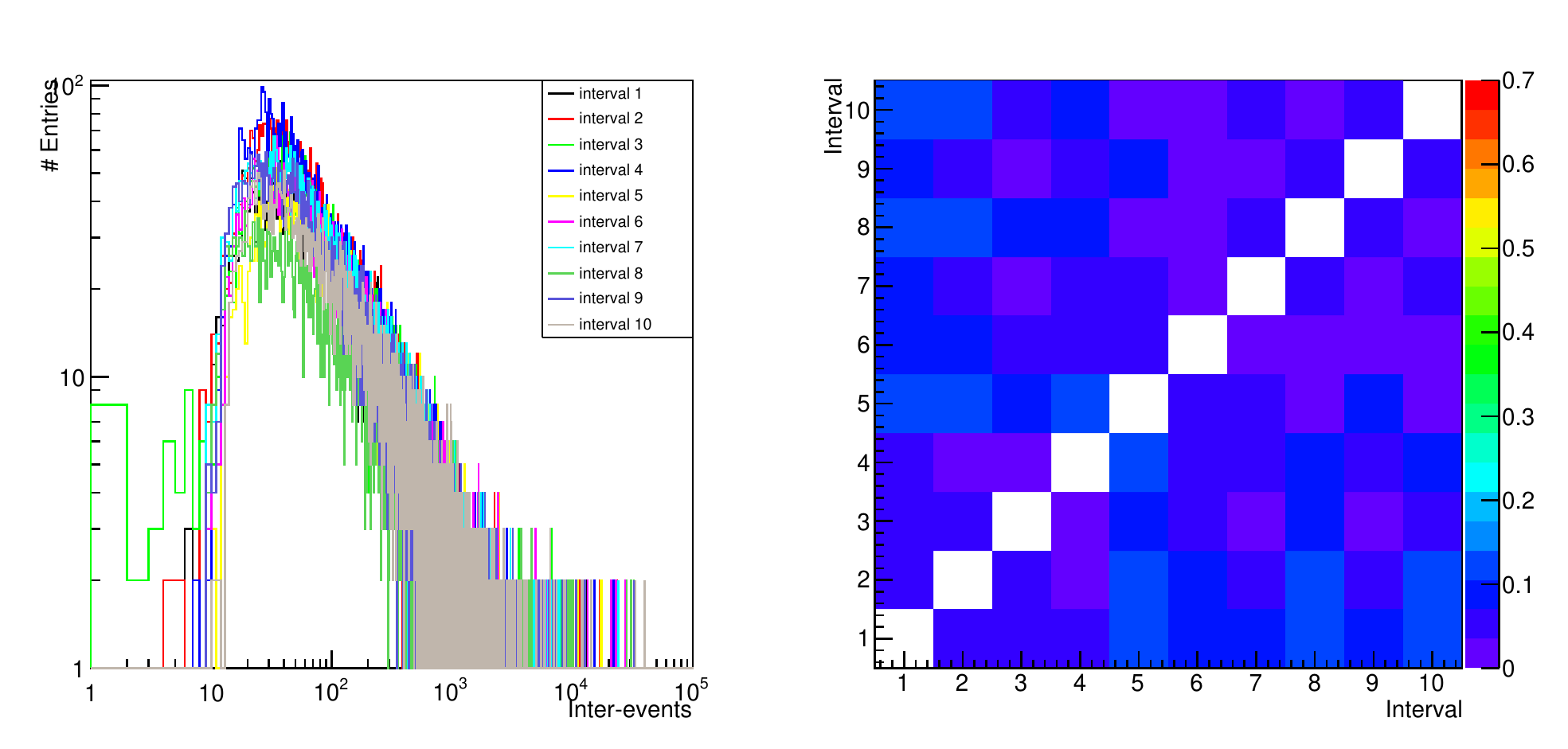}
\includegraphics[width=0.45 \textwidth]{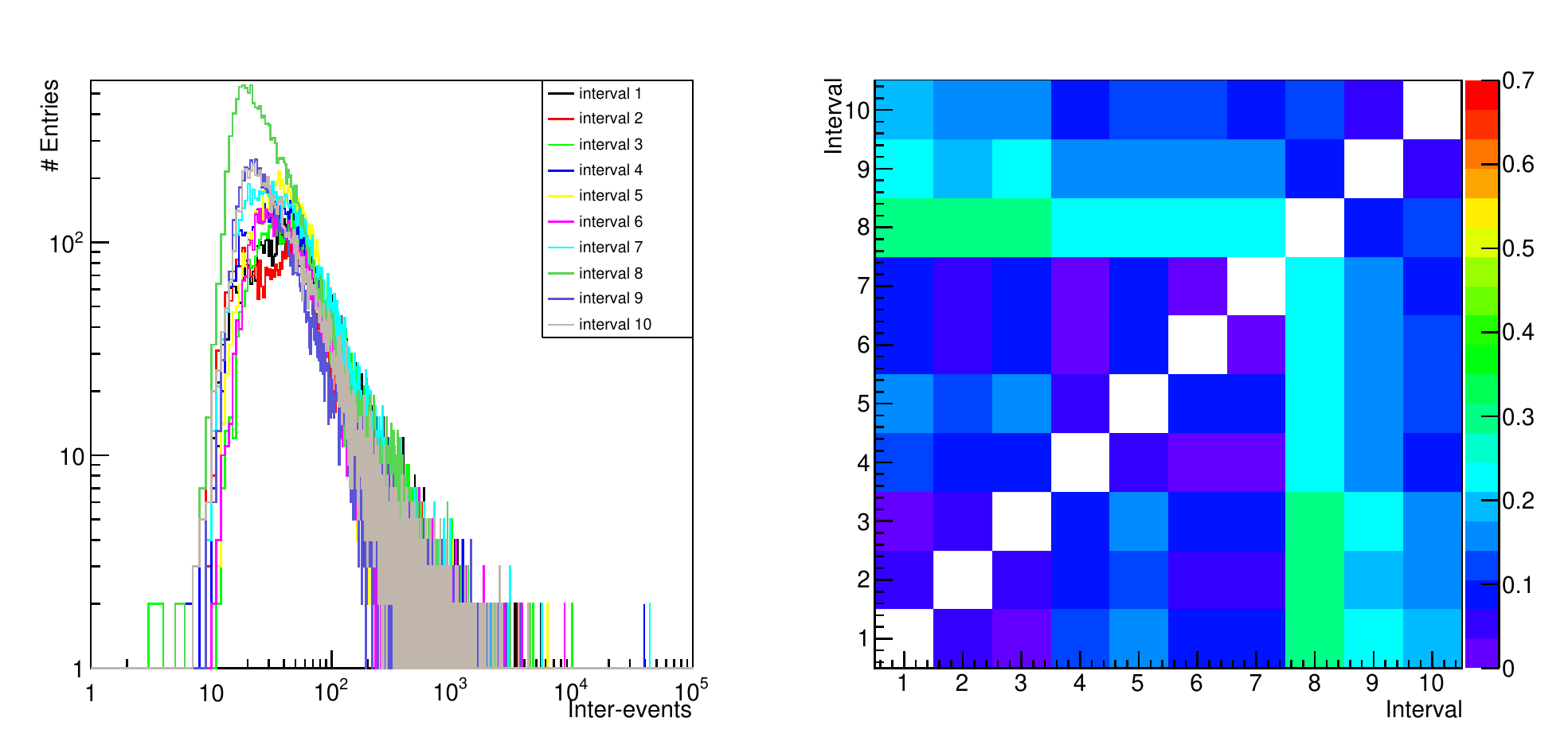}
\caption{Study of stationarity for ten time windows, covering the whole editing time span for editors $E_1$ (top left),  
$E_2$ (top right),  $E_3$ (bottom left) and  $E_4$ (bottom right). For each editor, in the left panel we show the inter-event distributions and in 
the right panel the results of the Kolmogorov-Smirnov test (comparison for all possible combinations of time intervals).}
\label{figks}
\end{center}
\end{figure}

In order to compare the ten probability distributions with one another, we used the Kolmogorov-Smirnov (KS) test. 
Some editors show, indeed, a lack of stationary behavior on the time scale of one year. However, there are some for which 
the probability distributions are very similar across the ten windows. As an example, we show in Fig.~\ref{figks} the results for four editors. 
For each editor, we show in the left panel the 
distribution of the inter-event times for each of the ten windows, while in the right panel we show the results of
the Kolmogorov-Smirnov test for all combinations of different time windows.  Both panels seem to indicate that the 
distributions shown on the left-hand side are stationary, while the distributions shown on the right-hand side become somewhat stationary after the 
fourth time window onwards. In the top row we show the results for editors $E_1$ and $E_2$ mentioned before, who have different 
degrees of behaviour towards stationarity. Nevertheless, our previous results stand for both of them, thus suggesting that 
the somewhat lack of stationarity of editor $E_2$ does not affect our previous results.

In table \ref{table:t1} we show the values of $B$ for each of the ten time windows, as well as the values for the whole data span. 
We can see that $B$ is high in each window, which seems to confirm the lack of correlation between burstiness and non-stationarity.

\begin{center}
\begin{table}
\caption{Values of burstiness for all data and in time windows, for editors $E_1$, $E_2$, $E_3$ and $E_4$.}
\begin{tabular}{|p{1.9cm}|p{1.3cm}|p{1.3cm}| |p{1.3cm}|p{1.3cm}| |p{1.3cm}|p{1.3cm}| |p{1.3cm}|p{1.3cm}|}         \hline
Data       &$B$ ($E_1$) & Length  &$B$ ($E_2$) & Length  &$B$ ($E_3$) & Length  &$B$ ($E_4$) & Length\\     \hline
all data  & 0.617  & 106824  & 0.912   & 267006  & 0.776 & 156876 & 0.850 & 130408     \\ \hline
window 1  & 0.638  & 9374    & 0.844   & 9955    & 0.846  & 11348 & 0.888 & 12529    \\
window 2  & 0.738  & 12023   & 0.800   & 15391   & 0.784  & 17427 & 0.829 & 9600    \\
window 3  & 0.632  & 11443   & 0.857   & 30343   & 0.702  & 10694 & 0.864 & 11967 \\ 
window 4  & 0.577  & 12912   & 0.898   & 26440   & 0.716  & 10492 & 0.785 & 12533     \\
window 5  & 0.549  & 10223   & 0.877   & 33079   & 0.755  & 16288 & 0.868 & 13641   \\
window 6  & 0.566  & 11316   & 0.896   & 21478   & 0.727  & 11687 & 0.806 & 10457 \\
window 7  & 0.568  & 11907   & 0.905   & 37167   & 0.736  & 19474 & 0.809 & 15273  \\
window 8  & 0.576  & 7170    & 0.890   & 24836   & 0.781  & 18496 & 0.840 & 21058  \\
window 9  & 0.574  & 10376   & 0.904   & 41526   & 0.759  & 19542 & 0.860 & 10839 \\
window 10 & 0.578  & 10080   & 0.958   & 26791   & 0.737  & 21428 & 0.829 & 12511 \\ \hline              
\end{tabular}
\label{table:t1}
 \end{table}
 
\end{center}

\section{Discussion}

In this work, we study bursty behavior in human activities and try to understand its origin. Using data
from Wikipedia editing, we have measured the average level of burstiness $B$ and
memory $M$ of editors, using standard indicators. Remarkably, our results fall outside the
region in the $(M,B)$ plane ascribed to human activity as proposed by
\cite{barabasi2008}, thus suggesting that such region should be enlarged.

We showed that the events are strongly conditioned by the circadian cycle of
$24$ hours, while the inter-events are stationary on the same time
scale. This suggests that burstiness is not a consequence
of the cyclic nature of human activity. Instead, it seems to be intrinsic to
mankind nature: before performing an action (make a phone call, write an email, edit
Wikipedia, etc) we must overcome a ``barrier", acting as a cost, which depends, among many other things,
on the time of day. However, once that ``barrier" has been crossed, the time taken by
that activity no longer depends on the time of day at which
we decided to perform it. Instead, there is a robust distribution of dedicated time in proportion to the different 
amount of activities that the individual is able to do at specific times. It could be related to some sort of queuing process, but
we prefer to see it as due to resource allocation (attention, time, energy),
which exhibits a broad distribution: shorter activities are more likely to be executed next than the
longer ones, which ultimately may be responsible for the bursty nature of human behavior. We verified that our results are independent of the effect 
of non-stationarity.

\vspace{1cm}
\textbf{Acknowledgments}

The work of Y.G., T.C. and R. L. presents research results of the Belgian
Network DYSCO (Dynamical Systems, Control, and Optimisation), funded by the
Interuniversity Attraction Poles Programme, initiated by the Belgian State, Science Policy Office.

\end{document}